\documentstyle[prd,aps,epsfig,draft]{revtex}
\begin{document}
\input  epsf.sty    
\twocolumn[\hsize\textwidth\columnwidth\hsize\csname
@twocolumnfalse\endcsname
%\documentstyle[aps,prl,twocolumn,floats,psfig]{revtex}
%\tighten
%\begin{document}
%
%\def\sqr#1#2{{\vcenter{\hrule height.3pt
%        \hbox{\vrule width.3pt height#2pt  \kern#1pt
%           \vrule width.3pt}  \hrule height.3pt}}}
%\def\square{\mathchoice{\sqr67\,}{\sqr67\,}\sqr{3}{3.5}\sqr{3}{3.5}}
%
%%%%%%%%%%%%%%%%%%%%%%%%%%%%%%%%%%%%%%%%%%%%%%%%%%%%%%%%%%%%%%%%%%%%%%%%%%

\sloppy

%\twocolumn
\draft

\title{ Hard Non-commutative Loops Resummation}
\author{Luca Griguolo$^1$ and Massimo Pietroni$^2$}
\address{{\it $^1$ Dipartimento di Fisica, Universit\`a di Parma,\\
INFN -- Gruppo Collegato di Parma,\\
Parco Area delle Scienze, 7/A, I-43100 Parma, Italy\\
and\\
$^2$ INFN -- Sezione di Padova,\\
Via F. Marzolo 8, I-35131 Padova, Italy}}
\maketitle

\begin{abstract}{The  non-commutative version of the euclidean $g^2 \phi^4$
 theory is considered. By using Wilsonian flow equations the 
ultraviolet renormalizability can be proved to all orders in
perturbation theory. On the other hand, the infrared sector cannot be 
treated perturbatively and 
requires a resummation of the leading divergences in the two-point function.
This is analogous  to what is done in the Hard Thermal Loops
 resummation of finite temperature field theory. 
Next-to-leading order corrections to the self-energy are computed, resulting 
in $O(g^3)$ contributions in the massless case, and $O(g^6 \log g^2)$ in the 
massive one.
 }
\end{abstract}
\pacs{PACS numbers: 11.10.Gh 11.10.Hi 11.25.-w\\
DFPD/01/TH/06,
UPRF-2001-02}

\vskip2pc]
\def\beqra{\begin{eqnarray}}

\def\eeqra{\end{eqnarray}}

\def\beq{\begin{equation}}

\def\eeq{\end{equation}}

\def\ds{\displaystyle}

\def\ts{\textstyle}

\def\ss{\scriptstyle}

\def\sss{\scriptscriptstyle}

\def\Vb{\bar{V}}

\def\phb{\bar{\phi}}

     \def\rhb{\bar{\rho}}

    \def\L{\Lambda}

    \def\T{\Theta}

\def\re#1{(\ref{#1})}

        \def\D{\Delta}

       \def\G{\Gamma}

\def\p{\partial}

\def\half{\mbox{\small$\frac{1}{2}$}}  

 \def\de{\delta}

%% macros to produce the symbols "less than or of order of"

%% and "greater than or of order of" %

\def\lta{\mathrel{\vcenter{\hbox{$<$}\nointerlineskip\hbox{$\sim$}}}}
\def\gta{\mathrel{\vcenter{\hbox{$>$}\nointerlineskip\hbox{$\sim$}}}}
\renewcommand{\Re}{\mathop{\mathrm{Re}}}
\renewcommand{\Im}{\mathop{\mathrm{Im}}}
\newcommand{\tr}{\mathop{\mathrm{tr}}}
\newcommand{\Tr}{\mathop{\mathrm{Tr}}}
%

%definitions to be erased for NPB version

%

\def\i{i}

 \def\f{f}

 \def\d{d}

 \def\e{e}

Quantum field theories on non-commutative spaces have been the subject of 
intense investigation in the recent past, mainly motivated  by 
their tight relation with string theories. Low energy 
excitations of a $D$-brane in a magnetic $B_{\mu\nu}$ background are indeed 
described by field theories with space non-commutativity \cite{seiwit}. In 
this limit the relevant description of dynamics is in term of massless 
open string states, while massive open string states and closed strings 
decouple: the full consistent string theory seems therefore truncated to the 
usual field theoretical degrees of freedom, suggesting the possibility that 
also the related quantum field theories are well defined. On the other hand 
their consistency is far from being obvious when 
examined from a purely field 
theoretical point of view: they are non-local (involving interactions with 
an arbitrarily 
high number of derivatives) and there is a new dimensionful 
parameter, other than the masses, taking into account the scale at which 
non-commutativity becomes relevant. 
%It is then natural to ask whether
%unitarity and renormalizability are implemented in these theories.
%For what concernes unitarity, it was shown in \cite{gomi} that
%when the non-commutativity involves space {\em and} time the perturbative 
%unitarity is indeed in trouble, whereas in the pure spatial case
%consistency with the Cutkoski's rules and unitary 
%time evolution has been checked \cite{gomi}.

It is then natural to ask whether these theories are renormalizable.
Contrary to 
early suggestions it was shown in ref.~\cite{filk} that infinities appear 
when perturbative computations are performed in non-commutative scalar 
theories. Moreover, an higly non-trivial mixture 
between ultraviolet (UV) and infra-red (IR) behaviors \cite{minwa}
makes  a proof of perturbative renormalization along the usual lines quite
cumbersome. Indeed, the one-loop  self-energy in the non-commutative version of
$g^2 \phi^4/4!$ theory gets a contribution of 
$O(g^2 M_{nc}^4/ p^2)$, where $M_{nc}$ is the scale of non-commutativity and 
$p$ the external momentum. This behavior is easily understood taking into 
account that the so called 
`non-planar' graphs are effectively cut-off in the UV  at a scale 
$O(M_{nc}^2/p)$
and that the scalar self-energy is quadratically divergent in the
commutative case~\cite{minwa}. 
When inserted in a higher order graph, the one-loop self-energy 
induces  IR-divergences even in the case of a massive theory. For instance,
the two-point function diverges quadratically in the IR for the massless 
theory at $O(g^4)$ and logarithmically for the massive one at $O(g^6)$, more 
tadpole insertions giving more and more IR-divergent behaviors.
A similar behavior was discovered for gauge theories \cite{suski}.

Due to this problematic `IR/UV connection' no complete calculation has been 
performed up to now in the scalar theory at next-to-leading order in 
perturbation theory. The possibility of absorbing UV divergences by means of
local counterterms has been discussed at two-loops in 
refs.~\cite{micu}, but no finite result could be obtained at that 
order due to the pathological behavior of  the integrals in the IR.

In this letter we will present the result of a resummation of 
the IR divergences
which allows a consistent computation of finite corrections beyond the
leading perturbative order. Before doing that, we will give the main 
lines of a proof of UV 
renormalizability to all orders in perturbation theory, which will be 
 presented in 
full detail in \cite{noi2}.

The need of a 
resummation has been realized by different people, and discussed for instance
in \cite{minwa,resu}. However, to our knowledge, 
no systematic approach has been formulated and therefore no 
explicit computation has been presented up to now.
In the resummed perturbative
expansion, the first correction to the leading $O(g^2)$ contribution to the 
self-energy of the massless theory arises at $O(g^3)$ instead of the naively 
expected $O(g^4)$. For the massive case, the resummation of all
the problematic diagrams first arising at $O(g^6)$, results in a 
$g^6 \log g^2$ behavior.

The pattern of the IR problem in the non-commutative scalar theory presents 
some remarkable similarity with that arising in finite temperature 
field theory~\cite{dolan}. 
In that case, the UV divergences of the new thermal contributions 
are cut-off by the 
temperature $T$, resulting in a $O(g^2 T^2)$ correction to the 
self-energy at one-loop.
The ratio between the one-loop amplitude and the tree-level one, 
$O(g^2 T^2/p^2)$ 
becomes of $O(1)$ for {\em soft} external momenta, $p\lta g T$, so that a 
resummation must be performed in order to define a sensible perturbative 
expansion. The program is accomplished by using a resummed propagator, in which
a `thermal mass' is included,  $(p^2+g^2 T^2/24)^{-1}$. Since the amplitude
which has been resummed receives contributions mainly by 
the {\em hard} ($p\sim T$) momentum part of the one-loop
integral, the resummed theory goes customarily under the name of 
`hard thermal loops'.
Analogously, we will use propagators in which one-loop amplitudes, 
dominated by momenta $O(M_{nc}^2/p)$, are resummed, and will consistently
call this procedure `hard non-commutative loops' resummation.

Before giving the details and results  of the resummation, we outline a 
formulation of the Wilsonian renormalization group (RG) {\`a} la Polchinski 
\cite{polchinski} for the non-commutative
scalar theory which will be presented in detail in a forthcoming publication.
By studying perturbatively the Wilsonian flow equation we are able to prove
UV renormalization to all orders. Then, turning to the IR
regime, we will show how a resummation procedure emerges quite
naturally in this context.

>From a field-theoretical point of view, the non-commutative version of 
$g^2 \phi^4/4!$ theory is again a scalar theory with the same tree-level 
propagator but a different vertex, which at tree-level is given by
\beqra
\label{FRULES}
\ds 
\Gamma^{(4)}(p_1,p_2,p_3,p_4)=&& g^2 h(p_1,p_2,p_3,p_4)\nonumber\\
\ds=&&
\frac{g^2}{3}\left[\cos(\frac{p_1\wedge p_2}{2})\cos(\frac{p_3\wedge p_4}{2})
\right.\nonumber\\ 
&&\left.+ ``(13)(24)''+
``(14)(23)'' 
\right]\,,
\eeqra
where $p\wedge q\equiv p_\mu \Theta_{\mu\nu}q_\nu$, $\Theta_{\mu\nu}$ being
the anti-symmetric matrix defining the commutation relations 
($[x_\mu,x_\nu]= \Theta_{\mu\nu}$).
As a consequence, the Wilsonian action and the RG equations can be derived 
in the same way
as in refs.~\cite{polchinski,bdm} 
-- they are actually {\em the same} equations, but with different 
boundary conditions. Our proof of perturbative UV renormalization parallels
quite closely the one given by Bonini, D'Attanasio and Marchesini for the
commutative case \cite{bdm},
so we only sketch here the main lines and stress only the differences 
emerging in the non-commutative theory.

A Wilsonian effective action can be defined, $\Gamma_{\L,\L_0}(\phi)$, as the
generating functional of 1PI Green functions obtained by integrating out loop
momenta $q$ such that $\L < q < \L_0$.  Our task is to prove that the double
limit $\L_0 \rightarrow \infty$ (UV renormalizability) and 
$\L\rightarrow 0$ (IR finitness) can be taken.

$\Gamma_{\L,\L_0}(\phi)$ and the Green function generated by 
$\phi$-derivating it, obey exact
evolution equations in $\L$.
There is a simple recipe to obtain the RG equation for any $2n$-point function;
{\it i)} write the 1-loop expression for $\Gamma^{(2n)}$ obtained by using
all the vertices up to $\Gamma^{(2n+2)}$, as if they were formally tree-level;
{\it ii)} promote  the tree-level 
vertices above to full, running, vertices,
$ \Gamma^{(2n)}\rightarrow \Gamma_{\L,\L_0}^{(2n)}$, and the tree-level 
propagator to the full, cut-off, propagator, 
\[
D_{\L,\L_0}(p)=
[(p^2+m^2) K(p;\L,\L_0)^{-1}+\Sigma_{\L,\L_0}(p)]^{-1}\,,
\]
where 
$\Sigma_{\L,\L_0}$ is the full, running, self-energy, and the cut-off function
$K(p;\L,\L_0)$ is equal to one in the interval $\L<p<\L_0$ and 
vanishes rapidly outside;
{\it iii)} take the derivative with respect to $\L$ everywhere in the 
$K$'s but not in the $\Sigma$'s or $\Gamma$'s. 

So, the evolution equation for {\em e.g.} the self-energy is given by
\[
%\label{RGE}
\ds \L\frac{\partial\;\;}{\partial \L}\Sigma_{\L, \L_0}(p) = \half
\int\frac{d^4 q}{(2\pi)^4}\frac{S_{\L, \L_0}(q)}{q^2+m^2} 
\Gamma^4_{\L, \L_0}(q,p,-p,-q)\;,
\]
where
\beqra
\label{KERNEL}
&&\ds \frac{S_{\L, \L_0}(q)}{q^2+m^2} \nonumber \\
\ds && \ds \equiv
\L\frac{\partial\;\;}
{\partial \L} 
\left.\frac{1}{(q^2+m^2) K(q;\L,\L_0)^{-1}+\Sigma_{\L^\prime,\L_0}(q)}
\right|_{\L^\prime=\L}\,\nonumber\\
&&\ds 
=\frac{1}{q^2+m^2} 
\frac{1}{\left[ 1+\frac{\Sigma_{\L, \L_0}(q)}{q^2+m^2}
K(q;\L,\L_0)  \right]^2} \L\frac{\partial\;\;}
{\partial \L}K(q;\L,\L_0)
\eeqra

The RG equations for higher point functions are obtained analogously, and they
together form an infinite system of coupled ordinary differential equations
which define the theory non-perturbatively.
The renormalization conditions are imposed by properly chosing the boundary
conditions. For the {\em relevant} vertices
\beqra
\label{RELOPS}
\gamma_2(\L)&&\equiv 
\left.\frac{d \Sigma_{\L, \L_0}(p)}{d p^2}\right|_{p^2=\mu^2},\;\;\;\;
\gamma_3(\L)\equiv \left.\Sigma_{\L, \L_0}(p)\right|_{p^2=\mu^2},\nonumber\\
\gamma_4(\L)&&\equiv \frac{
\Gamma^{4}_{\L, \L_0}(\bar{p}_1,\ldots,
\bar{p}_{4})}{h(\bar{p}_1,\ldots,
\bar{p}_{4})}
\eeqra
(where $\mu$ is the renormalization scale and the momenta $\bar{p}_i$ have 
been chosen such that 
$\bar{p}_i\cdot \bar{p}_j=\mu^2 (\delta_{ij}-\frac{1}{4})$) the boundary 
conditions are given at
the physical point $\L=0$,
\beq
\gamma_2(0)=1\,\,,\gamma_3(0)=\mu^2\,\,,\gamma_4(0)=g^2\,.
\eeq
The boundary conditions for all the other --{\em irrelevant}-- 
vertices (higher momentum derivatives in 
$\Sigma_{\L\L_0}$ and $\Gamma^{4}_{\L\L_0}$ and all the 
$\Gamma^{2n}_{\L\L_0}$'s with $n>2$)   are instead fixed at the UV,
$\L=\L_0$, where the irrelevant vertices can be set equal to zero.

With these boundary conditions, the relevant vertices at a generic $\L$ are 
given by integrals between $0$ and $\L$, whereas the irrelevant ones
are given by integrals between $\L$ and $\L_0$ \cite{bdm}. Then, one can 
disentangle the UV from the IR by taking the physical limit
 $\L_0 \to \infty$ and $\L\to 0$ in two successive steps.
The proof of UV renormalizability 
follows quite closely the well known one given by Polchinski 
 in the commutative case \cite{polchinski}, and  discussed further in 
\cite{bdm}. It exploits two remarkable features of the RG equations;
{\em i)} the momentum ordering, by which a given irrelevant coupling 
evaluated at cut-off $\L$ receives contributions only from loop momenta 
$\ge \L$, and {\em ii)} the one-loop structure of the exact equations, 
which make it possible to recover perturbation theory solving them iteratively.
As a result, the proof is extremely simple, as it is  based just on 
power counting arguments.
In the UV regime, the non-commutative and the commutative theories exhibit
essentially the same power counting.  Indeed, by chosing
$\L$ much larger than any physical scale of the theory, that is 
$\mu,m,M_{nc}\ll \L \ll \L_0$ the above mentioned
 $g^2 M_{nc}^4/q^2$ behavior of the 1-loop self-energy has not developed yet, 
as the momenta in the relevant integrals are bounded from below by 
$\L\gg M_{nc}$, and the one-loop self-energy is
subdominant with respect to the tree-level $q^2$ contribution in 
(\ref{KERNEL}). As a consequence, the $\L_0\to \infty$ limit 
can be shown to be finite at any perturbative order by a straightforward
translation of the arguments given in \cite{polchinski,bdm}.
 All the details of the proof will be given in a separate paper \cite{noi2}.

When the IR regime
comes under scrutiny, things change considerably. In \cite{bdm} the IR 
finitness of Green functions with non-exceptional external momenta 
({\it i.e} $p>O(\L)$) was proved for the commutative
 massless theory at any order in perturbation  theory. Crucial for that proof
is the fact that $\Sigma_{\L\L_0}$ is at most logarithmically divergent 
as $\L \rightarrow 0$ at any finite order in the expansion.
As we have repeatedly seen, this is not the case any more in the 
non-commutative case, where $\Sigma_{\L\L_0} \sim g^2 M_{nc}^4/\L^2$. 
Any perturbative computation is thus plagued by IR divergences 
which emerge sooner or later in the expansion in $g^2$. 

A quick look at the exact form of the RG evolution equations, 
and in particular at the kernel in eq. (\ref{KERNEL}), shows both what
the problem is and how a solution can be found. Namely, the more dangerous IR
divergences come out when --at any finite order in perturbation theory-- one 
expands the full-propagator appearing in the kernel in powers of 
$\Sigma_{\L\L_0}(q)/(q^2+m^2)$. Since $q\sim \L$ in the kernel, this ratio 
diverges as $\L^{-2}$ in the massive theory and as $\L^{-4}$ in the massless
one. It is then clear that any Green function 
-- even at non-exceptional momenta-- will be divergent at a sufficiently high 
order in $g^2$. 

The exact form of the RG kernel gives the solution as well. Since 
\mbox{$\Sigma_{\L\L_0}(q\sim\L)$}
 comes in the denominator, it is clear that the full
equations are indeed better behaved in the IR than any approximation
to them computed at any finite order in $g^2$. Actually, since the
effective mass explodes as $\L\rightarrow 0$, they are even better behaved
than those for the massive theory in the commutative case! 

It appears then clear
that the IR pathologies are just an artifact of the perturbative expansion,
which should disappear if this is properly reorganized.
To this end, one can still pursue the RG framework, splitting the full
two-point function as
\beqra
\label{SPLIT}
\ds 
\Gamma^{(2)}_{\L\L_0}(p) = &&(p^2+m^2)
K(p;\L,\L_0)^{-1}+\Sigma^{\mathrm LO}_{\L=0,\,\L_0}(p)\nonumber\\
&& + 
\Delta\Sigma_{\L,\L_0}(p)\,,
\eeqra
where $\Sigma^{\mathrm LO}_{\L=0,\,\L_0} = (g^2/24 \pi^2) 
[1-J_0(\L_0 \tilde{p}]/\tilde{p}^2$  is the leading IR contribution 
the one-loop self-energy (we 
have defined $\tilde{p}_\mu\equiv \Theta_{\mu\nu} p_\nu$, and $J_0$ is the 
Bessel function).
Eq.~(\ref{SPLIT})
 defines a new expansion in terms of $g^2$ and the new `tree-level'
 propagator $[(p^2+m^2)
K(p;\L,\L_0)^{-1}+\Sigma^{\mathrm LO}_{\L=0,\,\L_0}(p)]^{-1}$. 
All this can be done 
consistently in the Wilsonian RG framework, where the evolution equation for
$\Delta\Sigma_{\L,\L_0}(p)$ is easily obtained from that for the full
self-energy. Notice that, after resummation, the 
$\L_0\to \infty$ and $\L\to 0$ limits can be interchanged. 

In more common language, the resummation procedure simply amounts to adding 
and
subtracting the term 
\beq
\label{CT}
\frac{g^2}{48 \pi^2} \int 
\frac{d^4p}{(2\pi)^4} \phi(p)\frac{1}{\tilde{p}^2}\phi(-p)
\eeq
to the tree-level lagrangian (we have taken the $\L_0\to \infty$ limit),
 so as to get the resummed propagator provided
a new two-point `interaction' in eq.~(\ref{CT}) is 
consistently taken into account, in very
close analogy to what is done in the finite temperature theory case 
\cite{dolan}.

The interactions of the resummed theory give the 
Feynman rules in Fig.~1.
\begin{figure}[htb]
\epsfxsize=2.4 in \epsfbox{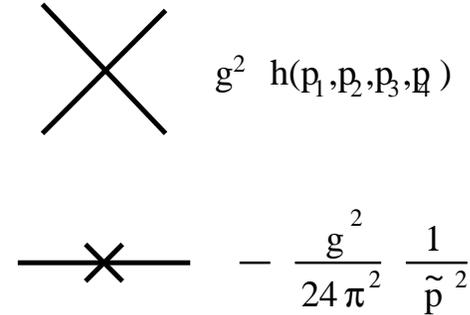}
\caption{The interaction vertices of the resummed theory}
\end{figure}
\noindent
Now we are ready to compute the next-to-leading order corrections to the 
self-energy, which are given by the two diagrams in Fig.~2, where the
resummed propagator runs into the loop (of course also the graph with the UV 
counterterms has to be included, which is not shown in the figure).
\begin{figure}[htb]
\leavevmode
\epsfxsize=1.8 in \epsfbox{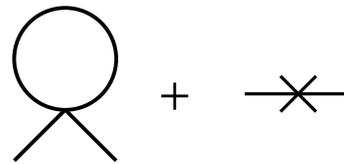}
\caption{The next-to-leading order contributions to the self-energy}
\end{figure}
\noindent
The tadpole diagram in the resummed theory gives
\beqra 
\Delta\Sigma(p)&& =
\frac{g^2}{6} \int\frac{d^4 q}{(2 \pi)^4} \frac{1}{q^2+m^2+\frac{g^2}{24\pi^2}
\frac{1}{\tilde{q}^2}} \left[2 + \cos(q\wedge p)\right] \nonumber\\
&&-
\frac{g^2}{24\pi^2}
\frac{1}{\tilde{p}^2} + ``{\mathrm UV c.t.}''\,. 
\eeqra
In the UV, the integral has the 
same structure as for the non-resummed theory, with a quadratically divergent
contribution from the `planar' diagrams and a finite one from the `non-planar'
ones, giving the $1/\tilde{p}^2$ term which is exactly cancelled by the 
new two-point interaction of the resummed theory.
In the IR, the planar and non-planar contributions sum up.
By writing 
\[\tilde{q}^2=\frac{1}{4}\Tr {\bf A} q^2 + q\cdot{\bf B} \cdot q\,,
\] 
where $A_{\mu\nu}=-\Theta_{\mu\rho}\Theta_{\rho\nu}$ and $B_{\mu\nu}$ is a 
traceless symmetric matrix. The symmetry of the integrand in the IR regime,
 selects  the $q^2$ term as the dominant contribution.

In the massless case ($m=0$) we find the following contribution from the 
`planar' graph
\beq
\label{MLP}
\Delta\Sigma^{m=0}_{\mathrm pl}(p)= -\frac{g^3}{96 \pi} M_{nc}^2 +O(g^5)\,,
\eeq
whereas from the `non-planar' one we get
 \beqra
\label{MLNP}
\Delta\Sigma^{m=0}_{\mathrm npl}(p)=&&\ds -\frac{g^3}{192 \pi} M_{nc}^2 
\nonumber \\
&& - \frac{g^4 M_{nc}^4 \tilde{p}^2}{1536 \pi^2} \left[
\log \frac{g^2 M_{nc}^4 \tilde{p}^4}{256}- {\mathrm const}\right]\,
\eeqra
for $M_{nc} \tilde{p} \ll 1$ and $
\Delta\Sigma^{m=0}_{\mathrm npl}(p)= O (g/M_{nc}^4 \tilde{p}^6)\,
$,
for $M_{nc} \tilde{p} \gg 1$, where we have defined 
$M_{nc}\equiv (6 \pi^2 \Tr {\bf A})^{-1/4}$.
In the massive case we get (planar + non-planar)
\beqra
\label{MASS}
\Delta\Sigma&&=\frac{g^2}{8 \pi^2}\left[m^2 \log m^2 -
\frac{g^2}{4}\frac{M_{nc}^4}{m^2}\right.\nonumber\\
\ds
&&\ds -\left.\frac{g^4}{8}\frac{M_{nc}^8}{m^6}
\left(\log\frac{g^4 M_{nc}^8}{m^8} +3\right) + O(g^6)\right]\,.
\eeqra
As one could expect, the non-analiticity in the coupling $g^2$ emerges 
at lower order in the massless case (where we find a $O((g^2)^{3/2})$ 
correction) compared to the massive one ($g^6 \log g^4$). This reflects
the fact that, in ordinary perturbation theory, 
the self-energy is IR divergent at $O(g^4)$ in the former 
case and  at $O(g^6)$ in the latter.

In computing the next-to-next-to-leading order in the resummed perturbative 
expansion one must
consistently take into account the two-point interaction in (\ref{CT}). 
Indeed, 
the two-loop graph for the resummed $m=0$ theory with one 
non-planar tadpole insertion (first graph in Fig.~3) 
gives a contribution of $O(g^3 M_{nc}^2)$, the 
same as the corrections computed above. It is only when the graph containing
the 
two-point interaction is added that the whole correction comes
out $O(g^5 M_{nc}^2)$.
\begin{figure}[htb]
\leavevmode
\epsfxsize=1.5 in \epsfbox{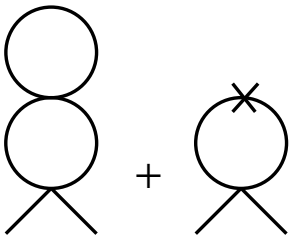}
\caption{The higher order corrections}
\end{figure}
\noindent
The $O(g^4)$ corrections that one gets at two-loop, coming from UV loop
momenta, cannot modify the $O(g^4 \log g^2)$ term in eq. (\ref{MLNP}).
The corrections computed above are really `perturbatively small' compared 
to the leading two-point function
 $p^2+g^2/24 \pi^2 \tilde{p}^2$ in any range of the momentum $p$. Indeed, for
large enough  momenta, the $\Delta \Sigma$ correction dominates over the 
$g^2/\tilde{p}^2$ term, but in that regime the tree-level $p^2$ term is 
leading. On the other hand in the IR the opposite happens, 
with $\Delta \Sigma$
never dominating over $g^2/\tilde{p}^2$.
Consequently, no 
 tachyonic  behavior can be induced by the next-to-leading order 
corrections.

%The proof of UV renormalizability sketched in the first part of 
%this letter can be repeated for the resummed theory, as  resummation 
%only modifies the IR sector. While in the unresummed case the two limits 
%$\L_0 \to \infty$ and $\L\to 0$ do not commute, they do for the resummed case.
%Indeed, consider for instance the two-loop graph on the left in Fig. 3 for 
%the massive theory. If the
%propagators are not resummed the integral in the IR goes as
%\beq
%\int_\L \frac{d\lambda\;  \lambda^3}{(\lambda^2+m^2)^2} \times
%\left[ \frac{M_{NC}^4}{\lambda^2} \;\; (\L_0 \to \infty), \;\; \;
%\L_0^2 \;\;(\L_0 {\mathrm fixed})\right],
%\label{IRUV}
%\eeq
%thus giving contributions $O(M_{NC}^4/m^2)$ and $O(\L_0^2 \log m^2)$ 
%respectively, 
%when $\L\to 0$. This 
%is the effect of IR/UV mixing from the one-loop self-energy. On the other 
%hand, in the resummed theory the denominator in (\ref{IRUV}) is replaced by 
%$(\lambda^2+m^2+g^2 M_{NC}^4/\lambda^2)^2$, thus giving in both cases vanishing
%IR contributions, respectively $O(\L^6/M_{NC}^4)$ and 
%$O(\L_0^2 \L^8/M_{NC}^{8})$.

The resummed propagator was interpreted in ref. \cite{minwa} as originating 
from some high-energy degrees of freedom which, when integrated out, leave a 
commutative scalar theory with modified dispersion relations. In the RG 
language, the contributions to the two-- and four--point functions  induced
by non-commutativity may be analogously seen as high-energy 
boundary conditions of an otherwise commutative  
theory valid up to some energy $\bar{\L} \ll M_{NC}$ \cite{noi2}.

It would be interesting to know if a resummation can be
performed  in gauge theories on the same spirit of this paper. 
Modifying only the two-point function is not a gauge invariant operation, so 
that an hypotetical resummation must necessarily involve all Green 
functions, as it is the case in hard-thermal-loop resummed QCD \cite{bp}.
\acknowledgments

We thank Marisa Bonini for useful discussions.

\end{document}